\documentclass[twocolumn,showpacs]{revtex4}
\usepackage{graphicx}
\usepackage{dcolumn}
\usepackage{bm}

\usepackage{amsmath}

\begin{document}

\title{Theory of Interferometric Photon-Correlation Measurements: Differentiating Coherent from Chaotic Light}

\author{A.\ Lebreton, I.\ Abram, R.\ Braive, I.\ Sagnes, I.\ Robert-Philip$^*$, and A.\ Beveratos}

\address{Laboratoire de Photonique et Nanostructures LPN-CNRS UPR-20, Route de Nozay, 91460 Marcoussis, France\\$^*$Corresponding author: isabelle.robert@lpn.cnrs.fr }

\begin{abstract}
Interferometric photon-correlation measurements, which correspond to the second-order intensity cross-correlations between the two output ports of an unbalanced Michelson interferometer, are sensitive to both amplitude and phase fluctuations of an incoming beam of light. 
Here, we present the theoretical framework behind these measurements and show that they can be used to unambiguously differentiate a coherent wave undergoing dynamical amplitude and phase fluctuations from a chaotic state of light. 
This technique may thus be used to characterize the output of nanolasers and monitor the onset of coherent emission.
\end{abstract}

\pacs{42.50.Ar, 
42.60.Mi, 
05.40.Ca, 
42.55.Sa 
}

\maketitle

\section{Introduction}

The laser threshold highlights a radical change in the mode of operation of the device. 
When the laser operates at powers below threshold, the emission of the excited dipoles constituting the active medium is understood to be spontaneous (such that on average there is less than one photon in the cavity) and the laser output is incoherent.
The incoherent field is described as a statistical ensemble of waves emitted at random times with random phases and displays chaotic (thermal) photon statistics \cite{Glauber, Loudon}. 
On the other hand, at powers above threshold, the emission is stimulated (as the number of photons in the cavity exceeds one) and the laser output is coherent.
The state of the field is described by a quantum mechanical coherent state, which is usually represented in phase-space as a rotating vector (albeit with quantum uncertainty) called a ``phasor'', and displays Poisson photon statistics  \cite{Loudon}.

The coherence of the laser output is usually evaluated by measuring two types of autocorrelation functions.
The first-order (field) autocorrelation function is usually measured by means of an interferometer that introduces a delay $\tau$ between its two arms, and is defined as
\begin{equation}
g^{(1)}(\tau)= \frac{\langle\hat{E}^*(t+\tau)\hat{E}(t) \rangle_{t}}{\langle\hat{E}^*(t)\hat{E}(t) \rangle_{t}}
\end{equation}
where $\hat{E}^*(t)/\hat{E}(t)$ is the positive/negative frequency part of the electric field operator at time $t$, and $\langle...\rangle_t$ denotes a quantum mechanical and statistical average, as well as an average over time $t$. 
For an ergodic stationary field, the statistical averaging is equivalent to the time-average, so that in that case the autocorrelation function may be evaluated at an arbitrary time, e.g.\ at $t=0$.
The first-order autocorrelation function is essentially the Fourier transform of the field spectrum, and thus for a Lorentzian spectrum it corresponds to a decaying exponential,
\begin{equation}
g^{(1)}(\tau)= e^{-i\omega\tau}e^{-|\tau|/\tau_c}
\end{equation}
where $\omega$ is the frequency of the field and $\tau_c$ is its coherence time.
Both chaotic and coherent states of light with Lorentzian spectra give first-order correlation functions of the form of Eq.\ (\theequation).
Measuremant of $\tau_c$ permits nevetheless a quantitative distinction between the chaotic or coherent output of a laser, as the value of $\tau_c$ becomes longer above threshold, reflecting the narrowing of the Schawlow-Townes linewidth of the laser as a function of pump power.
However, as Glauber pointed out in his seminal papers on the theory of coherence: ``Coherence does not require monochromaticity. Coherent fields can be generated with arbitrary spectra'' \cite{Glauber}, implying that a measurement of $g^{(1)}$ is insufficient for identifying a coherent state.

The second-order (intensity) autocorrelation function, defined as
\begin{equation}
g^{(2)}(\tau)= \frac{\langle\hat{E}^*(t)\hat{E}^*(t+\tau)\hat{E}(t+\tau)\hat{E}(t) \rangle_{t}}{\langle\hat{E}^*(t)\hat{E}(t) \rangle_{t}^2}
\end{equation}
is usually measured in a Hanbury Brown and Twiss photon-counting experiment, as the probability of detecting two photons separated by a time-difference $\tau$.
The second-order autocorrelation function measures essentially the time-distribution of the relative intensity fluctuations and its value at zero time-difference ($\tau=0$), gives the variance ($\sigma^2$) of the relative intensity fluctuations of the field through the expression $\sigma^2=g^{(2)}(0)-1$.
The value of $g^{(2)}(0)$ is often used to characterize the coherence of the laser output, as it is equal to 2 for chaotic light and equal to 1 for a stable coherent wave with no amplitude fluctuations.

In conventional (large) lasers, when the pump power is increased across the threshold, there is an abrupt transition in the value of the zero time-difference second-order intensity autocorrelation from $g^{(2)}(0)=2$ to $g^{(2)}(0)=1$, a feature that has led to the identification of this transition with the threshold \cite{Rice1994}.
However, in nanolasers the transition between the two values of $g^{(2)}(0)$ as a function of input power is very gradual, even when an abrupt change in output power clearly identifies a threshold.
This feature is often interpreted as indicating that the output of nanolasers is not fully coherent above threshold, but includes incoherent spontaneous emission \cite{Ulrich2007, Choi2007, Hofmann2000}.

This interpretation rests on the assumption that the value of $g^{(2)}(0)$ has a direct (bijective) relationship with coherence.
However, while the value $g^{(2)}(0)=1$ can only be obtained for a coherent stable wave, the values $g^{(2)}(0)>1$ do not necessarily correspond to the presence of incoherent chaotic light in the laser output. 
They may also arise from a coherent quantum state undergoing dynamical amplitude fluctuations.
In this context, we note that in conventional large lasers the value $g^{(2)}(0)=1$ arises because their gain (i.e.\ the number of excited dipoles) is clamped above threshold, so that the output field presents no amplitude fluctuations if the pumping rate is stable.
In a laser, the coherent field undergoes phase fluctuations (corresponding to the Schawlow-Townes phase diffusion) and these give rise to its finite spectral width, while preserving the value $g^{(2)}(0)=1$ since the gain remains clamped.
On the other hand, if the gain were not clamped but fluctuated in time, it would produce amplitude (intensity) fluctuations in the laser output, giving rise to $g^{(2)}(0)>1$.
Thus, the value of $g^{(2)}(0)$ alone is not sufficient for distinguishing between chaotic and coherent light.

In a recent publication, we presented a novel experimental technique, consisting of interferometric photon correlation measurements, which gives qualitatively different results when applied to chaotic and to coherent light and can thus unambiguously discriminate between these two types of states of the electromagnetic field \cite{Lebreton2013}. 
This technique, a variant of which was initially developed to study the dynamical fluctuations in the fluorescence spectrum of single molecules \cite{Brokmann2006, Coolen2007}, can measure the correlation times of both phase and amplitude fluctuations of an incoming beam.
It can thus distinguish a coherent field from a statistical distribution of randomly-phased waves (chaotic field) as in the latter case the two types of fluctuations have a fixed relationship. 
By use of this technique, we showed that the output of a nanoscale laser is indeed a coherent state undergoing amplitude fluctuations, and not a chaotic statistical ensemble of spontaneously-emitted waves.
We thus confirmed earlier publications in which we presented evidence that the fluctuations in the output of a nanolaser is not due to the presence of incoherent spontaneous emission in the output of the laser, but is the result of the laser's relaxation oscillations excited by the ``discretization noise'' associated with the small (and integer) number of photons at threshold \cite{Elvira2011}.

In this paper, we develop the theory of interferometric photon correlation experiments and examine the corresponding correlation function for three types of states of the electromagnetic field: (1) chaotic light (2) stable coherent light and (3) coherent light with amplitude fluctuations.
We show that the interferometric cross-correlation function for chaotic light is qualitatively different from that of coherent light, thus identifying a signature for coherent light.
We thus show that even though chaotic light and amplitude-modulated coherent light may have similar first- and second-order correlation functions, they may be distinguished by their interferometric cross-correlation functions.

\section{Interferometric photon correlation measurements}

\begin{figure}[h]
   \begin{center}
   \begin{tabular}{c}
   \includegraphics[width=7.6cm]{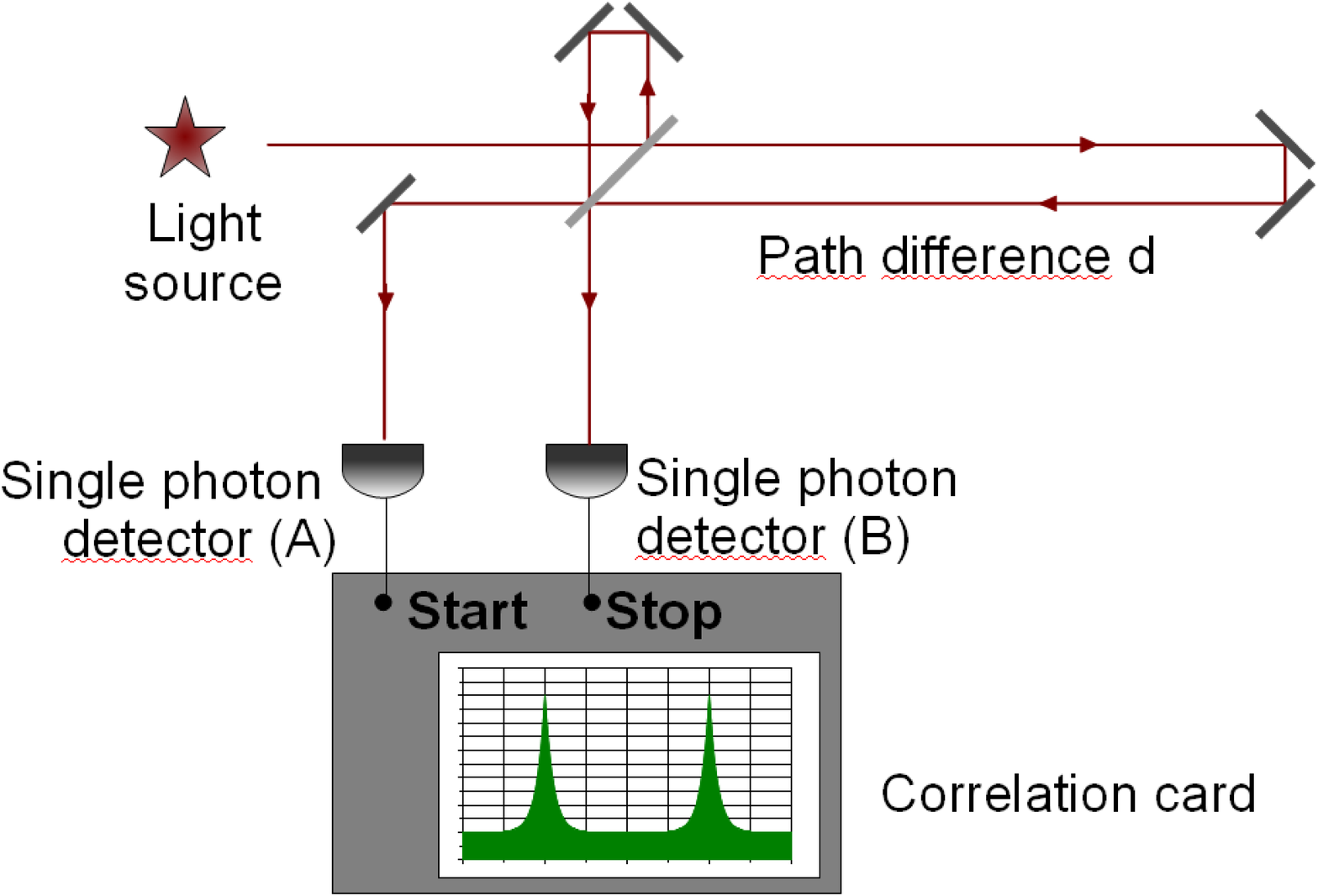}
   \end{tabular}
   \end{center}
   \caption[example]
   { \label{Fig:Michelson} Experimental setup for interferometric photon correlation measurements.
   Light enters an unbalanced Michelson interferometer. 
   The pathlength difference between the two arms of the interferometer $d$ is dithered over several wavelengths to average out the interference fringe contrast. 
   Two single-photon detectors at the two output ports of the interferometer ($A/B$) measure the cross-correlation in the photon arrival times.}
\end{figure}

We consider a light beam of central frequency $\omega$ and coherence time $\tau_{c}$ (for simplicity we take the spectrum to be Lorentzian), entering an unbalanced Michelson interferometer (see Fig.\ \ref{Fig:Michelson}) in which one arm is longer by a distance $d$, thus introducing a delay of $d/c$ (where $c$ is the speed of light) with respect to the other. 
The electric field operators at the two output ports $A/B$ of the interferometer may be written in terms of the input field operators as
\begin{eqnarray}
\hat{E}^*_A(t) & = & E_0^* \frac{ a^\dagger (t+d/c)- a^\dagger (t)}{\sqrt2}
 \nonumber \\ 
\hat{E}^*_B(t) & = & E_0^* \frac{ a^\dagger (t+d/c)+ a^\dagger (t)}{\sqrt2}
\label{eq:EAB}
\end{eqnarray}
where $E^*_0$ and $a^\dagger (t)$ are, respectively, the vacuum field and the creation operator for the input mode.
An equivalent pair of equations holds for $\hat{E}_A/\hat{E}_B$ and the annihilation operators.
Using these relations, the intensity coming out at each port for a stationary input field, may be written as 
\begin{eqnarray}
\langle \hat{E}^*_A(t) E_A(t)\rangle& = & \frac{I_1}{2} (1-2 \Re [g^{(1)}(d/c)]) \nonumber \\
\langle \hat{E}^*_B(t) E_B(t)\rangle& = & \frac{I_1}{2} (1+2 \Re [g^{(1)}(d/c)])
\end{eqnarray}
where $I_1=|E_0|^2 \langle a^\dagger a\rangle$ is the input intensity and $\Re[x]$ is the real part of the complex number $x$.

The quantity that is measured in these experiments is the second-order intensity cross-correlation function between the two output ports of the interferometer, which may be written as:
\begin{equation}
g^{(2X)}(\tau,d/c) = \langle\hat{E}^*_A(0)\hat{E}^*_B(\tau)\hat{E}_B(\tau)\hat{E}_A(0) \rangle
\label{eq:g2x}
\end{equation}
where $\tau$ is the time-difference between the photon detection events in ports $A$ and $B$, while the normalization of $g^{(2X)}$ through a denominator equal to the product of the two mean output intensities is implicit.
Expressing $g^{(2X)}$ in terms of the input field operators through Eqs.\ (\ref{eq:EAB}), 16 terms are obtained. 
Among these terms, those that involve an unequal number of creation and annihilation operators delayed by $d/c$ (for example, the term $\langle a^\dagger (0) a^\dagger (d/c+\tau) a(\tau) a(0) \rangle$) oscillate like $e^{i\omega d/c}$ or $e^{2i\omega d/c}$ (or their complex conjugates) and thus average out to zero when the interferometer arm difference $d$ is dithered over a distance of a few wavelengths. 
The six remaining terms are:
\begin{eqnarray}
g^{(2X)}(\tau,d/c) =  \frac{1}{4} \langle && a^\dagger (d/c) a^\dagger (d/c+\tau) a(d/c+\tau) a(d/c) 
 \nonumber \\ 
 &+& a^\dagger (0) a^\dagger (\tau) a(\tau) a(0) 
 \nonumber \\ 
 &+& a^\dagger (0) a^\dagger (d/c+\tau) a(d/c+\tau) a(0)
 \nonumber \\ 
 &+& a^\dagger (d/c) a^\dagger (\tau) a(\tau) a(d/c)
 \nonumber \\ 
 &-& a^\dagger (d/c) a^\dagger (\tau) a(d/c + \tau) a(0)
 \nonumber \\ 
&-& a^\dagger (0) a^\dagger (d/c+\tau) a(\tau) a(d/c) \rangle
\label{eq:6terms}
\end{eqnarray}

The first two terms correspond each to $g^{(2)}(\tau)$, the ``standard'' second-order autocorrelation function of the input field, and arise from the two photons propagating like particles along the same interferometer arm, while the third and fourth terms correspond respectively to $g^{(2)}(\tau+d/c)$ and $g^{(2)}(\tau-d/c)$ (that is to the second-order autocorrelation function displaced by $\tau=-d/c$ and $\tau=d/c$) and arise from the two photons propagating like particles along different interferometer arms. 
These four terms are sensitive only to the intensity fluctuations of the field.
The last two terms, which have a negative sign, describe the fluctuations of the field when it undergoes interference by propagating through both arms of the interferometer.
It is these terms, which will be called hereafter ``interference terms'', that permits us to differentiate chaotic from coherent light, as they are sensitive to both amplitude and phase fluctuations.
They correspond to the conditional probability that a constructive interference of the two paths with difference $d$ results in the detection of a second photon at $t=\tau$, given that a constructive interference at $t=0$ led to the detection of a first photon.
Eq. (\ref{eq:6terms}) may thus be re-written as
\begin{widetext}
\begin{equation}
g^{(2X)}(\tau,d/c) =  \frac{1}{4} \left \{ 2 g^{(2)}(\tau)+g^{(2)}(d/c+\tau)+g^{(2)}(d/c-\tau) - 2 \Re [ g^{(2)}(d/c,\tau,d/c+\tau,0)] \right \}
\label{eq:g2X}
\end{equation}
\end{widetext}
where 
$g^{(2)}(t_4,t_3,t_2,t_1)= \langle a^\dagger(t_4)  a^\dagger(t_3) a(t_2) a(t_1)\rangle$ is the four-time second-order correlation function of the input field that corresponds to the ``interference terms''.

We note that when the interferometer is balanced ($d=0$), the interference terms reduce to the standard intensity autocorrelation function and thus the interferometric cross-correlation function reduces to 
\begin{equation}
g^{(2X)}(\tau,0) =  \frac{1}{2}  g^{(2)}(\tau)
\end{equation}
where the factor 1/2 arises from the averaging over several interference fringes due to the dithering: 
When the interferometer is exactly balanced, light comes out only at port $B$ and the cross-correlation is equal to zero; however, when the interferometer is between two fringes light comes out from both ports, corresponding to a cross-correlation of one, giving an average of 1/2 upon dithering.
However, when the interferometer is unbalanced, introducing a delay longer than the coherence time of the incoming radiation $d/c> \tau_c$, the interferometric cross-correlation function deviates from Eq.\ (\theequation) and its features depend on the nature of the incoming radiation.
In the next few sections, we will examine the features of $g^{(2X)}(\tau,d/c)$ for a chaotic field, described as a statistical ensemble of waves and for a coherent field, described either as a single stable wave with phase fluctuations, or as a wave with both phase and amplitude fluctuations.

\section{Chaotic field}

We consider the electromagnetic field produced by a laser pumped below threshold, in which $N$ excited dipoles emit spontaneously and independenty, each one photon. 
This field may be described as a statistical ensemble of $N$ randomly-phased waves, so that its electric field may be written as \cite{Loudon}
\begin{equation}
E^*(t)= E^*_0 \frac{e^{i \omega t}}{\sqrt{N}}\sum_j^N e^{i \phi_j(t)}
\end{equation}
where the phase $\phi_j$, associated with the wave emanating from emitter $j$, undergoes random jumps with correlation time $\tau_c$, according to the equation of motion,
\begin{equation}
\dot{\phi}_j=f_j
\end{equation}
where randomly fluctuating frequency $f_j$ is a Langevin noise source obeying the standard correlation relations,
\begin{eqnarray}
\langle f_j(t)\rangle& = & 0 \nonumber \\
\langle f_j(t_2)f_k(t_1)\rangle& = & \frac{2}{\tau_c} \delta_{jk} \delta(t_2-t_1)
\end{eqnarray}
where $\delta_{jk}$ is the Kronecker $\delta$-symbol, while $\delta(t_2-t_1)$ is the Dirac $\delta$-function.

\begin{figure}[h]
   \begin{center}
   \begin{tabular}{c}
   \includegraphics[width=7.6cm]{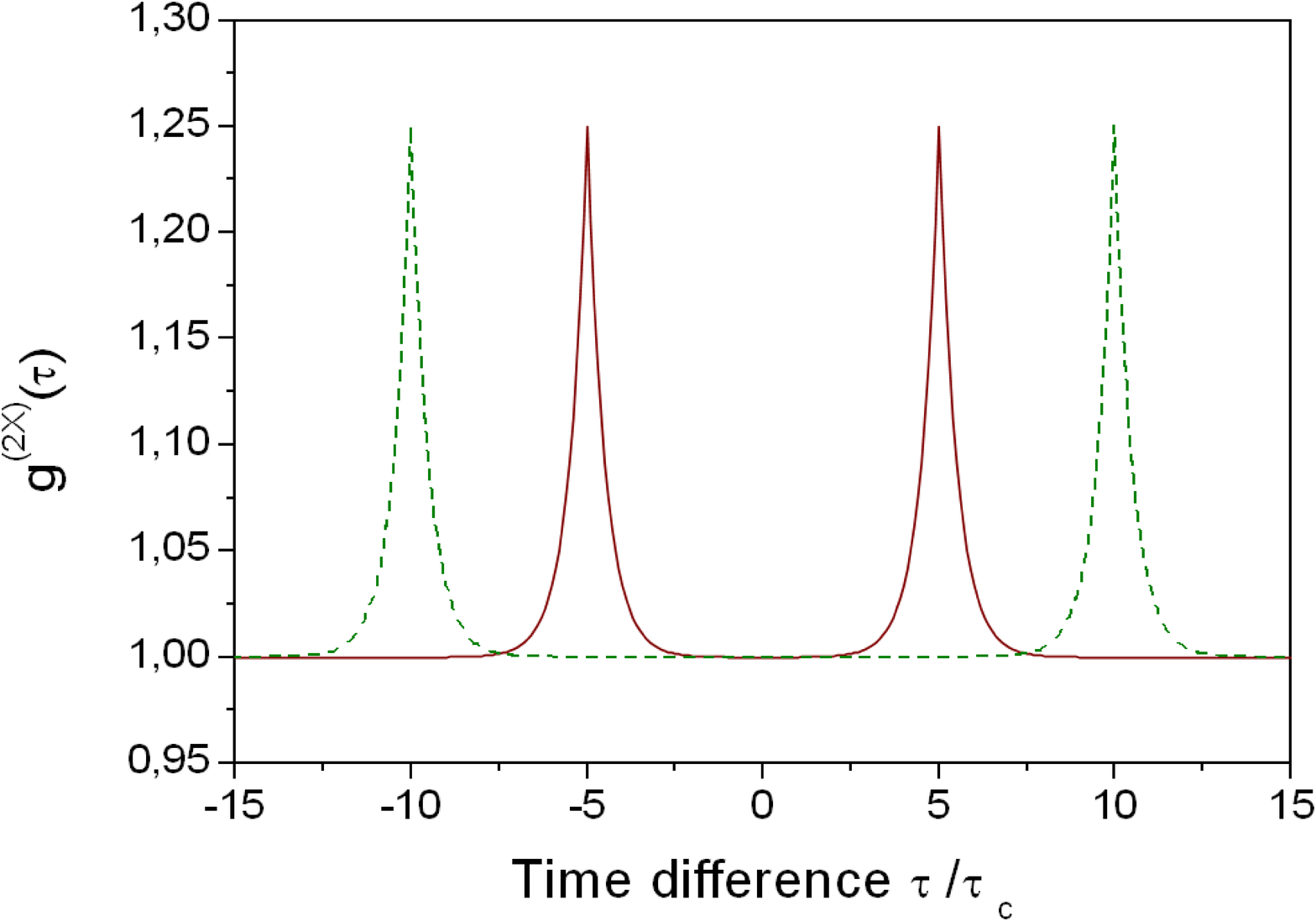}
   \end{tabular}
   \end{center}
   \caption[example]
   { \label{Fig:chaotic} 
   Interferometric second-order intensity cross-correlation function for a chaotic field $g^{(2X)}_{chaotic}(\tau, d/c)$ at the output ports of an unbalanced Michelson interferometer with pathlength difference $d=5 c \tau_c$ (continuous red curve) and $d=10 c \tau_c$ (dashed green curve).}
\end{figure}

The first-order correlation function of this field may be calculated as
\begin{eqnarray}
g^{(1)}(\tau)&=& \frac{e^{-i \omega \tau}}{N}\sum_{j,k}^N \langle e^{i \phi_j(\tau)-i \phi_k(0)} \rangle \nonumber \\
&=& e^{-i \omega \tau}e^{ - |\tau|/\tau_c}
\end{eqnarray}
while the second-order correlation function is:
\begin{eqnarray}
g^{(2)}(\tau)&=& \frac{1}{N^2}\sum_{j,k,\ell,m}^N \langle e^{i \phi_j(0)+i \phi_k(\tau)-i \phi_\ell(\tau)-i \phi_m(0)} \rangle \nonumber \\
&=& 1+ |g^{(1)}(\tau)|^2= 1+e^{- 2|\tau|/\tau_c}
\end{eqnarray}
It corresponds to a peak at $\tau=0$ with a value of $g^{(2)}(0)=2$ and decaying to $g^{(2)}(\infty)=1$ exponentially at a rate $2/\tau_c$.
The interference terms in $g^{(2X)}$ (Eq.\ \ref{eq:g2X}) are:
\begin{widetext}
\begin{eqnarray}
g^{(2)}(d/c,\tau,d/c+\tau,0) &=& \frac{1}{N^2}\sum_{j,k,\ell,m}^N \langle e^{i \phi_j(d/c)+i \phi_k(\tau)-i \phi_\ell(d/c+\tau)-i \phi_m(0)} \rangle \nonumber \\
&=& e^{- 2|d|/c\tau_c}+e^{- 2|\tau|/\tau_c}
\end{eqnarray}
\end{widetext}
so that the interferometric second-order intensity cross-correlation function $g^{(2X)}$ for a chaotic field is,
\begin{widetext}
\begin{equation}
g^{(2X)}_{chaotic}(\tau, d/c)= 1 + \frac14 e^{- 2|d/c+\tau|/\tau_c} +\frac14 e^{- 2|d/c-\tau|/\tau_c} - \frac12 e^{- 2|d|/c\tau_c}
\end{equation}
\end{widetext}
Thus, it presents two peaks at $\tau=\pm d/c$ (see Fig.\ \ref{Fig:chaotic}), corresponding to a pair of initially bunched photons traveling each on a different arm of the interferometer. Its value at $\tau=0$ is $g^{(2)}(0,d/c)=1$, independent of the path-length difference $d$ (averaged over several interference fringes), since the two photons of two interfering elementary randomly-phased waves have equal probability of coming out in either one of the two output ports.

\section{Stable coherent field}

\begin{figure}[h]
   \begin{center}
   \begin{tabular}{c}
   \includegraphics[width=7.6cm]{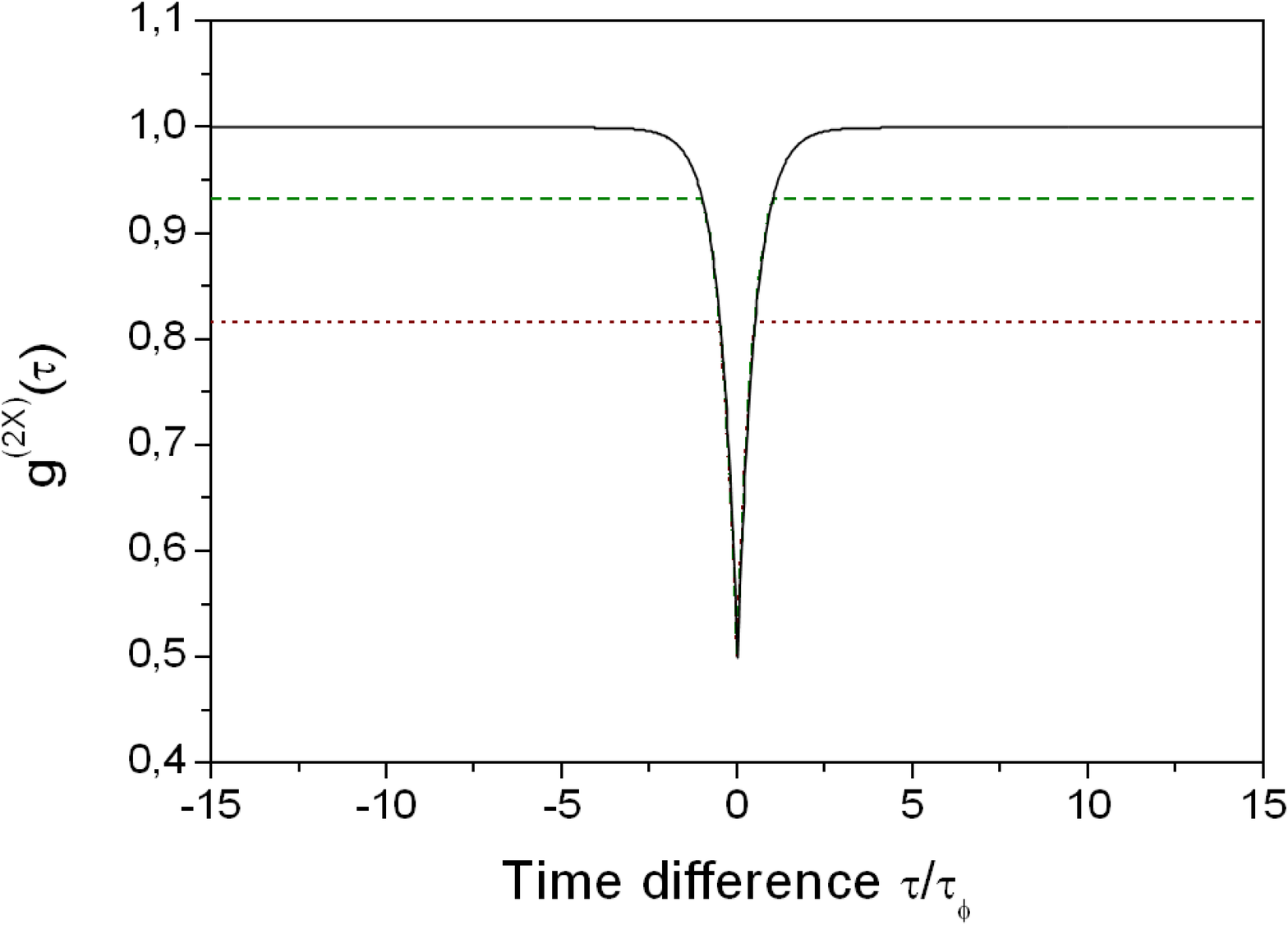}
   \end{tabular}
   \end{center}
   \caption[example]
   {\label{Fig:coherent} 
   Interferometric second-order intensity cross-correlation function for a stable coherent wave $g^{(2X)}_{coh}(\tau, d/c)$ at the output ports of an unbalanced Michelson interferometer with pathlength difference $d= 0.5 c \tau_\phi$ (dotted red curve), $d= c \tau_\phi$ (dashed green curve) and $d \gg c \tau_c$ (continuous black curve).}
\end{figure}

We consider the light produced by a conventional laser operating above threshold.
The state of the electromagnetic field can be described as a coherent quantum mechanical state that consists of a superposition of all number-states of the lasing mode, such that it presents a non-zero and oscillating expectation value for the electric field operator.
The expectation value of the electric field thus corresponds to a classical oscillating stable wave.
At the same time, the laser field undergoes phase diffusion describable by the Schawlow-Townes process, so that the electric field may be written as
\begin{equation}
E^*(t)=E^*_1 e^{i \omega t+i \phi(t)}
\end{equation}
where $E_1^*=E^*_0\sqrt{\langle a^\dagger a \rangle}$ is the coherent amplitude of the electric field, while the phase $\phi$ undergoes random jumps with correlation time given by the Schawlow-Townes phase diffusion time $\tau_\phi$ according to 
\begin{equation}
\dot{\phi}=f_\phi
\end{equation}
where $f_\phi$ is a randomly fluctuating Langevin noise source obeying
\begin{eqnarray}
\langle f_\phi(t)\rangle& = & 0 \nonumber \\
\langle f_\phi(t_2)f_\phi(t_1)\rangle& = & \frac{2}{\tau_\phi} \delta(t_2-t_1)
\end{eqnarray}
The first-order correlation function of this field is
\begin{eqnarray}
g^{(1)}(\tau)&=& {e^{-i \omega \tau}}\langle e^{i \phi(\tau)-i \phi(0)}\rangle \nonumber \\
&=& e^{-i \omega \tau}e^{ - |\tau|/\tau_\phi}
\end{eqnarray}
the second-order correlation function is
\begin{equation}
g^{(2)}(\tau)= 1
\end{equation}
reflecting the lack of photon bunching in a coherent state.
The interference terms in $g^{(2X)}$ (Eq.\ \ref{eq:g2X}) may be calculated as:
\begin{eqnarray}
g^{(2)}(d/c,\tau,d/c+\tau,0) &=& \langle e^{i \phi(d/c)+i \phi(\tau)-i \phi(d/c+\tau)-i \phi(0)}\rangle \nonumber \\
&=& e^{- 2 m /\tau_\phi}
\end{eqnarray}
where $m=\min (|d/c|,|\tau|)$ is the smaller of the two values $|d/c|$ or $|\tau|$.
Thus the interferometric second-order intensity cross-correlation function for a stable coherent field undergoing phase diffusion is
\begin{equation}
g^{(2X)}_{coh}(\tau, d/c)= 1- \frac12 e^{- 2 m/\tau_\phi}
\end{equation}
It consists of a flat baseline at $g^{(2X)}(\infty, d/c)= 1- \frac12 e^{- 2 d/(c\tau_\phi)}$, in which there is a dip at $\tau=0$ going down to the value of $g^{(2X)}(0, d/c)=0.5$ (see Fig.\ \ref{Fig:coherent}). 
This dip is the signature of a coherent state and is most pronounced when the interferometer is strongly unbalanced ($d \gg c \tau_\phi$) as in this case the baseline assumes its highest value $g^{(2X)}(\infty, d/c)=1$. 
Its full-width measured at the value of $1-1/(2e)=0.816$ gives directly the correlation time of the random phase jumps, $\tau_\phi$.
Its value at $\tau = 0$, significantly below 1, indicates that if a first photon is detected in one particular output port (say, in port $A$), a second photon arriving at times $\tau \ll \tau_\phi$ will also exit by the same port since for those short times the coherent state retains the memory of its phase and constructive interference will occur for the same output port. At long times however, $\tau \gg \tau_\phi$, the second photon may exit from either port.

\section{Coherent field with amplitude fluctuations}

A coherent field undergoing strong amplitude fluctuations in addition to Schawlow-Townes phase diffusion may be described by a simple phenomenological model consisting of a wave modulated in amplitude (or in intensity) by an external signal whose frequency $f_{mod}$ varies randomly, following the correlation relations
\begin{eqnarray}
\langle f_{mod}(t)\rangle& = & 0 \nonumber \\
\langle f_{mod}(t_2)f_{mod}(t_1)\rangle& = & \frac{1}{2\tau_{int}} \delta(t_2-t_1)
\end{eqnarray}
where $\tau_{int}$ is the correlation time of the intensity fluctuations.
We may further assume for simplicity that the amplitude and phase fluctuations are uncorrelated, and that the amplitude modulation depth is large, so that the unmodulated part of the carrier wave may be ignored.
The electric field may thus be written as, 
\begin{equation}
E^*(t)=\sqrt2 E^*_1 \cos(\chi (t)) e^{i\omega t+i\phi(t)}
\end{equation}
where the modulation phase $\chi(t)$ is driven by the randomly-varying frequency according to
\begin{equation}
\dot{\chi} (t)=f_{mod}
\end{equation}

The first-order correlation function is
\begin{eqnarray}
g^{(1)}(\tau) &=& 2\langle \cos(\chi (\tau)) \cos(\chi (0)) \rangle e^{i\omega \tau} \langle e^{i\phi(\tau)- i\phi(0)} \rangle \nonumber \\
&=&  e^{i\omega \tau} e^{-|\tau|/(4\tau_{int})} e^{-|\tau|/\tau_\phi} =  e^{i\omega \tau} e^{-|\tau|/\tau_c}
\end{eqnarray}
where the correlation time $\tau_c$ includes both phase and amplitude (intensity) fluctuations,
\begin{equation}
\frac{1}{\tau_c} = \frac{1}{\tau_\phi} + \frac{1}{4\tau_{int}}
\end{equation}

The second-order correlation function for a wave undergoing amplitude fluctuations may be calculated as
\begin{eqnarray}
g^{(2)}(\tau) &=& 4\langle \cos^2(\chi (\tau)) \cos^2(\chi (0)) \rangle \nonumber \\
&=& 1+\frac12e^{-|\tau|/\tau_{int}}
\end{eqnarray}
It thus resembles the second-order correlation function for a chaotic field with a peak at $\tau=0$, but unlike the chaotic field that peak is not related to the first-order correlation function $g^{(1)}(\tau)$ and the spectrum of the field.
It is nevertheless related to its power spectrum, so that Eq.\ (\theequation) may also be expressed in terms of the Fourier transform of the Relative Intensity Noise (RIN) spectrum, $FT[RIN]$, as
\begin{equation}
g^{(2)}(\tau) =1+FT[RIN]
\end{equation}
which, in turn, is defined in terms of the spectral density of the intensity fluctuations, $S_{\delta I}$, as
\begin{equation}
RIN=\frac{2 S_{\delta I}}{I_1^2}
\end{equation}
The interference terms in $g^{(2X)}$ (Eq.\ \ref{eq:g2X}) may be calculated as:
\begin{widetext}
\begin{eqnarray}
g^{(2)}(d/c,\tau,d/c+\tau,0) &=& 4\langle \cos(\chi (d/c)) \cos(\chi (\tau)) \cos(\chi (d/c+\tau)) \cos(\chi (0))\rangle \langle e^{i \phi(d/c)+i \phi(\tau)-i \phi(d/c+\tau)-i \phi(0)}\rangle \nonumber \\
&=& \left ( 1+\frac12 e^{- M /\tau_{int}} \right ) e^{-2 m /\tau_c}
\end{eqnarray}
\end{widetext}
where $M=\max (|d/c|,|\tau|)$ is the larger of the two values $|d/c|$ or $|\tau|$, 
so that the $g^{(2X)}$ may be written as,
\begin{widetext}
\begin{equation}
g^{(2X)}_{amp-fluct}(\tau, d/c) = 1 + 
\left [ \frac14e^{-\tau/\tau_{int}}  - \frac12 \left ( 1+\frac12 e^{- M /\tau_{int}} \right ) e^{-2 m /\tau_c} \right ]
+ \frac18 e^{-|d/c+\tau|/\tau_{int}} + \frac18 e^{-|d/c-\tau|/\tau_{int}}
\label{eq:amp}
\end{equation}
\end{widetext}
where the terms that give rise to features in the central part of the curve (i.e. near $\tau \approx 0$) have been grouped inside square brackets [...] for clarity.

\begin{figure}[h]
   \begin{center}
   \begin{tabular}{c}
   \includegraphics[width=7.6cm]{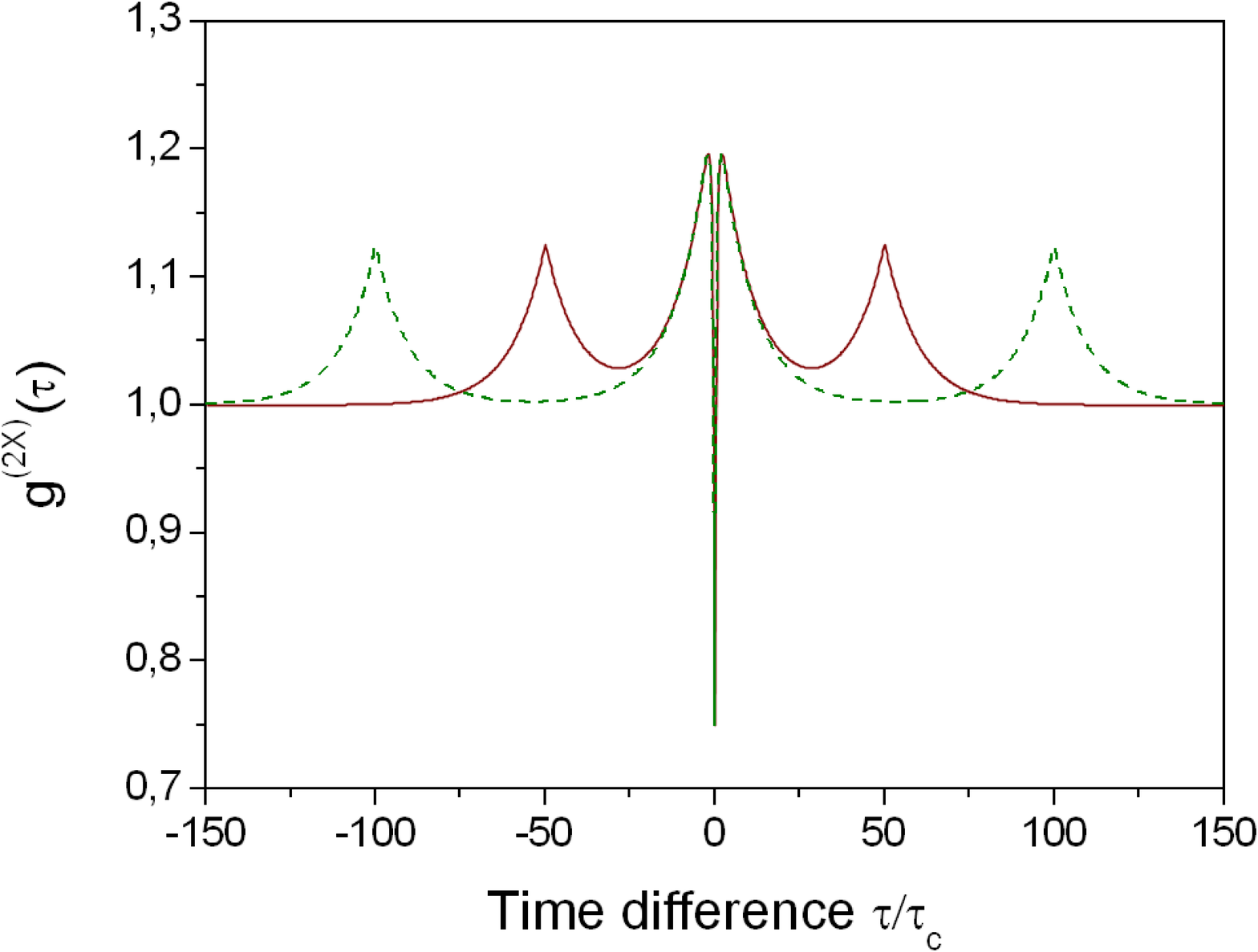}
   \end{tabular}
   \end{center}
   \caption[example]
   { \label{Fig:amp} 
   Interferometric second-order intensity cross-correlation function for a coherent field with intensity fluctuations $g^{(2X)}_{amp-fluct}(\tau, d/c)$ at the output ports of an unbalanced Michelson interferometer with pathlength difference $d= 50 c \tau_c = 5 c \tau_{int}$ (continuous red curve), $d=100 c \tau_c= 10 c \tau_{int}$ (dashed green curve). 
      The characteristic time of the intensity fluctuations is taken to be $\tau_{int}=10 \tau_c$.}
\end{figure}

The interferometric second-order intensity cross-correlation function for a coherent field undergoing amplitude modulation at a randomly varying frequency (see Fig\ \ref{Fig:amp}), bears the signature of a coherent state which is a dip at $\tau \approx 0$ going down to $g^{(2X)}_{amp-fluct}(0,d/c)=0.75$ and displaying a width of $2/\tau_c$. 
However, in contrast to the case of a stable coherent wave, the presence of amplitude fluctuations cause the central part of $g^{(2X)}_{amp-fluct}(\tau, d/c)$ to rise to a broad peak of width $1/\tau_{int}$ inside which the sharp coherence dip appears, since $2/\tau_c <1/\tau_{int}$.
Thus, the central part of $g^{(2X)}_{amp-fluct}(\tau, d/c)$ gives both the coherence time of the laser and the correlation time of its amplitude fluctuations. 
Away from the central region, $g^{(2X)}_{amp-fluct}(\tau, d/c)$ displays two peaks at $\tau = \pm d/c$ corresponding to two displaced replicas of the autocorrelation function, as is also the case for the chaotic field.

\begin{figure}[h]
   \begin{center}
   \begin{tabular}{c}
   \includegraphics[width=7.6cm]{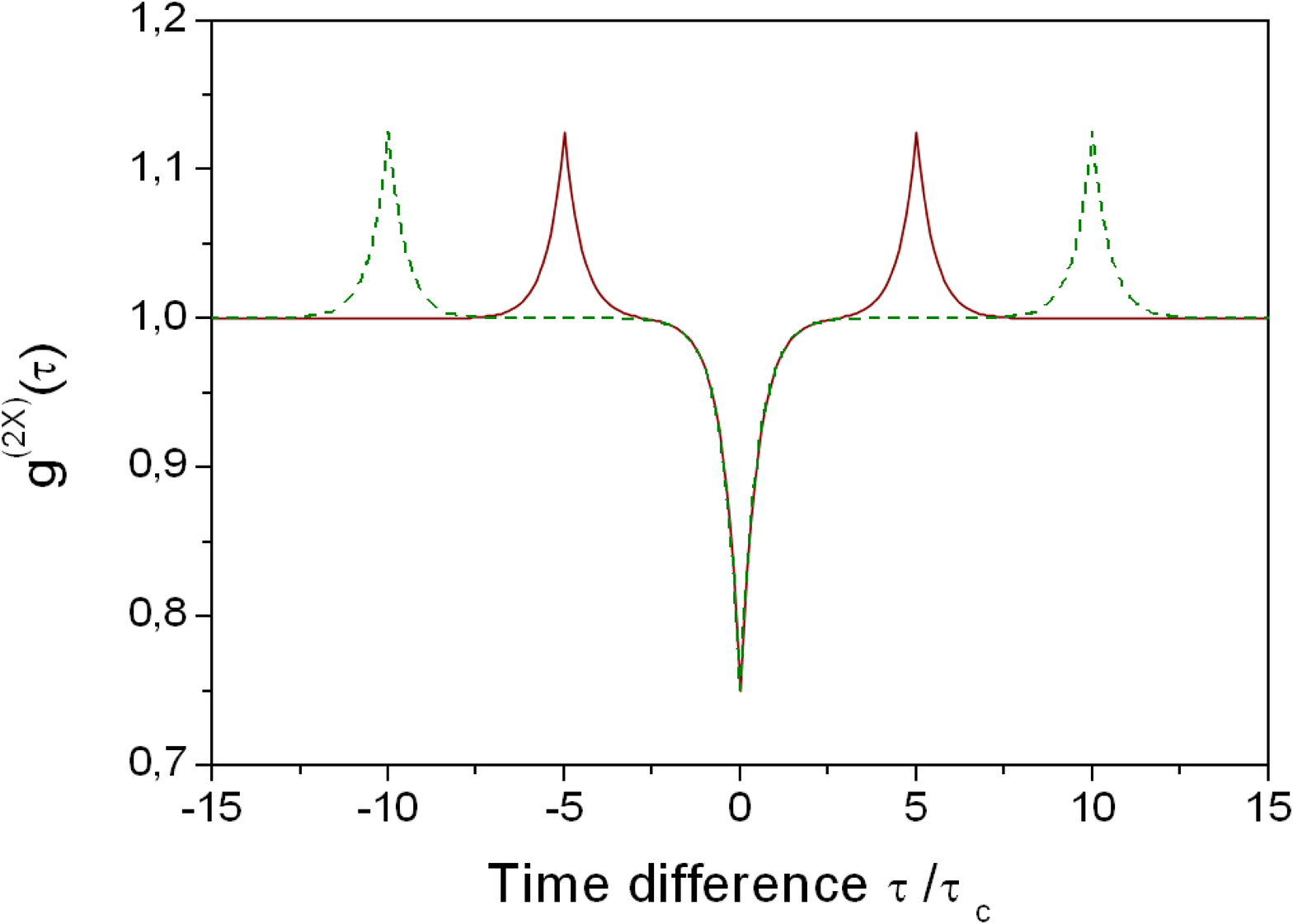}
   \end{tabular}
   \end{center}
   \caption[example]
   { \label{Fig:mix} 
   Interferometric second-order intensity cross-correlation function for a 50-50 ``mixture'' of coherent and chaotic light $g^{(2X)}_{mix}(d/c,\tau)$ at the output ports of an unbalanced Michelson interferometer with pathlength difference $d= 5 c \tau_c$ (continuous red curve), $d=10 c \tau_c$ (dashed green curve).}
\end{figure}

It should be noted that a ``mixture'' consisting of a fraction $x$ coherent and $(1-x)$ chaotic light displays an interferometric cross-correlation function of the form 
\begin{widetext}
\begin{equation}
g^{(2X)}_{mix}(\tau,d/c) = 1  - \left [ \frac{x}{2} e^{-2 m /\tau_\phi} \right ]
+ \frac{1-x}{4} e^{- 2|d/c+\tau|/\tau_c} +\frac{1-x}{4} e^{- 2|d/c-\tau|/\tau_c} - \frac{1-x}{2} e^{- 2|d|/c\tau_c}
\end{equation}
\end{widetext}
where the term that gives rise to a feature in the central part of the curve (i.e. near $\tau \approx 0$) is inside square brackets [...] for comparison with Eq.\ (\ref{eq:amp}).
As can be seen on Fig.\ \ref{Fig:mix}, for a coherent-chaotic ``mixture'' the coherence dip appears on a flat baseline whose value is 1, for $d \gg c \tau_c$. There is no peak rising above the value of 1 in the central part of this curve, in contrast to the case for coherent light with amplitude fluctuations (compare with Fig.\ \ref{Fig:amp}).
Thus, the presence of the broad central peak together with a dip in the interferometric cross-correlation function (arising from the terms in square brackets in Eq.\ (\ref{eq:amp})) permits to unequivocally discriminate a coherent field undergoing amplitude fluctuations from a ``mixture'' of coherent and chaotic light, even though both types of fields display similar second-order correlation functions $g^{(2)}(\tau)$.

\section{Summary and conclusion}

Interferometric photon correlation measurements, in which the beamsplitter of a standard Hanbury Brown and Twiss setup is replaced by an unbalanced Michelson interferometer, are sensitive to the dynamical fluctuations of the incoming field and thus can discriminate between a field fluctuating in time and a statistical ensemble of randomly-phased waves. 
They can therefore be used to differentiate coherent (but fluctuating) light from chaotic light and thus pinpoint the onset of coherence of a laser.

The unbalanced Michelson interferometer, which superposes two time-separated parts of the field, embodies Glauber's pioneering view of coherence: ``In physical optics the term is used to denote a tendency of two values of the field at distantly separated points or at greatly separated times to take on correlated values...The coherence conditions restrict randomness of the fields rather than their bandwidth'' \cite{Glauber}.
Alternatively, the measurement of the second-order intensity cross-correlation function at the output ports of the interferometer is equivalent to a photon-counting self-homodyne detection in the time domain, with the signal reflected back from the long arm corresponding to the local oscillator.

The second-order cross-correlation function at the two outputs of an unbalanced Michelson interferometer, $g^{(2X)}(\tau, d/c)$, is qualitatively different for chaotic and for coherent fields.
While $g^{(2X)}_{chaotic}(\tau, d/c)$ for a chaotic field consists of two peaks corresponding to two replicas of $g^{(2)}$ displaced to $\tau = \pm d/c$ and is flat around $\tau \approx 0$ assuming the value $g^{(2X)}(0,d/c)=1$, the second-order cross-correlation function for a coherent field with amplitude fluctuations, in addition to the two replicas of $g^{(2)}$ displaced to $\tau = \pm d/c$, it displays a dip at $\tau = 0$ whose width is given by twice the inverse of the coherence time of the field, superimposed on a broad peak that corresponds to $g^{(2)}(\tau)$.
These features, consisting of a sharp dip dug into a broad peak, constitute the signature of a coherent field undergoing amplitude fluctuations and are absent for a chaotic field.
Thus, inspection of the central part of $g^{(2X)}(\tau, d/c)$ (i.e.\ around $\tau \approx 0$) permits an unequivocal discrimination between a coherent and a partially chaotic field.

Acknowledgment: The authors acknowledge financial support from the C'NANO APOLLON project and from the French National Research Agency (ANR) through the Nanoscience and Nanotechnology Program (project NATIF ANR-09-NANO-P103-36).

\end{document}